\documentclass[sigconf]{acmart}

\usepackage{amsmath,amsfonts}
\usepackage{graphicx}
\usepackage{subcaption}
\graphicspath{ {/plots/without_edge_filtering/} }
\usepackage{url}
\usepackage{xcolor}
\usepackage{dirtytalk}
\usepackage{enumitem}
\usepackage{graphicx}
\usepackage{caption}
\usepackage{subcaption}
\usepackage{float}

\usepackage{xcolor}

\usepackage{url}
\usepackage{xcolor}

\setcopyright{acmcopyright}
\copyrightyear{2020}
\acmYear{2020}

\begin{document}

\title{Analysis of Twitter and YouTube during USelections 2020}


\author{Alexander Shevtsov}
\email{shevtsov@ics.forth.gr}
\affiliation{
  \institution{FORTH-ICS}
  \institution{Computer Science Dept., U.~of Crete}
}
\author{Maria Oikonomidou}
\email{mareco@ics.forth.gr}
\affiliation{
  \institution{FORTH-ICS}
  \institution{Computer Science Dept., U.~of Crete}
}
\author{Despoina Antonakaki}
\email{despoina@ics.forth.gr}
\affiliation{
  \institution{FORTH-ICS}
}
\author{Polyvios Pratikakis}
\email{polyvios@ics.forth.gr}
\affiliation{
  \institution{FORTH-ICS}
  \institution{Computer Science Dept., U.~of Crete}
}

\author{Sotiris  Ioannidis}
\email{sotiris@ece.tuc.gr}
\affiliation{
  \institution{Technical University of Crete}
  \institution{FORTH-ICS}
}

%

%




\renewcommand{\shortauthors}{Shevtsov, Oikonomidou, et al.}

\begin{abstract}

The presidential elections in the United States on 3 November 2020 have caused extensive
discussions on social media. A part of the content on US elections is organic,
coming from users discussing their opinions of the candidates, political positions,
or relevant content presented on television. Another significant part of the
content generated originates from organized campaigns, both official and by astroturfing.

In this study, we obtain approximately 17.5M tweets containing 3M users, based
on prevalent hashtags related to US election 2020, as well as the related YouTube links,
contained in the Twitter dataset, likes, dislikes and comments of the videos and conduct volume,
sentiment and graph analysis on the communities formed.

Particularly, we study the daily traffic per prevalent hashtags, plot the retweet graph from July to September 2020, show how its main connected component becomes denser in the period closer to the elections and highlight the two main entities ('Biden' and 'Trump').  
Additionally, we gather the related YouTube links contained in the previous dataset
and perform sentiment analysis. The results on sentiment analysis on the Twitter
corpus and the YouTube metadata gathered, show the positive and negative sentiment
for the two entities throughout this period.
The results of sentiment analysis indicate that 45.7\% express positive sentiment
towards Trump in Twitter and 33.8\% positive sentiment towards Biden, while 14.55\% of
users express positive sentiment in YouTube metadata gathered towards Trump and 8.7\%
positive sentiment towards Biden. 
Our analysis fill the gap between the connection of offline events and their consequences
in social media by monitoring important events in real world and measuring public
volume and sentiment before and after the event in social media.

\end{abstract}




\maketitle

\section{Introduction}

Today social networks have a very important role in online social discourse and especially during pre-elections period. OSNs can either be used productively to perform dissemination, communication or administration in elections, like in \cite{howard2018algorithms}. The content posted by the users can represent their political belief or is either used to comment, be sarcastic or express negative opinion towards a political party or ideology. Twitter and YouTube while constitute two of the most popular online social networks attracting million of users daily, capture a very important proportion of this online discourse. 

Through analysis of this online discourse we can discover the main tendencies and preference of the electorate, generate patters that can distinguish users' favouritism towards an ideology or a specific political party, study the sentiment prevailed towards the political parties or even predict the outcome of the elections. Analyzing the sentiment is a necessary step towards any of these directions. We are not focusing on the prediction of the electoral outcome, or sarcasm detection, but rather exploring the sentiment towards the political parties in Twitter and YouTube, as well. 

Sentiment analysis have been extensively studied in Twitter, usually studying a specific event. Through sentiment analysis we can visualize the variation of sentiment of the electorate during a political event [\cite{wang2012system, diakopoulos2010characterizing}], a company event [\cite{daniel2017company}], product reviews [\cite{mukherjee2012feature}] or model the public mood and emotion and connect tweets' sentiment features with fluctuations with real events [\cite{bollen2009modeling}]. The main task of these works is to predict elections, classify the electorate and distinguish posts towards one political party or ideology, although is has been addressed in many works that Twitter is not suitable for elections prediction [\cite{gayo2013meta, gayo2011limits}].

Sentiment in these analyses is represented by a variable with values like 'Positive', 'Negative' and 'Neutral', or even more specific ('Happy','Angry', etc.). Each word in the corpus can be assigned with more than one sentiment ('Positive' and 'Negative'). Other metrics that can be measured in this analysis, are 'subjectivity' and 'polarity', where the first one is defined as the ratio of 'positive' and 'negative' tweets to 'neutral' tweets, while the second is defined as the ratio of 'Positive' to 'Negative' tweets. There is a plethora of works of sentiment analysis in Twitter [\cite{martinez2014sentiment, giachanou2016like, go2009twitter, mittal2012stock, saif2012alleviating}]. 

Sentiment analysis usually requires 'text normalization', an initial preprocessing of the corpus in order to extract the lexical features that can significantly affect the performance [\cite{kolchyna2015twitter, pak2010twitter, jianqiang2017comparison}]. The steps of the preprocessing include tokenization, expansion of abbreviations and removal of stop words (URLs, mentions etc.).

In this study  we obtain the most popular hashtags around the US elections and gather a dataset of 7.5M tweets, for a period of three months. We extract 16.642 unique YouTube videos contained in this dataset as well as their metadata (likes, comments, authors etc.). Initially, we perform a volume analysis and an association of the diverse features of the YouTube videos. The next step is to identify two main entities (‘Trump’ and ‘Biden’) in our corpus, in order to perform sentiment analysis. Next, we study the retweet graph in six different time points in our dataset, from July to September 2020 and highlight the two main entities (’Biden’ and ’Trump’). The final section is the sentiment analysis by utilizing Vader sentiment analysis model and we show the higher positive sentiment towards Donald Trump in Twitter (45.7\%) and in YouTube (14.55\%) in comparison to the positive sentiment expressed towards Joe Biden, by the users in Twitter(33.8\%) and YouTube (8.7\%).

\subsection{Background}
Some methods incorporate the use of Twitter features, like emoticons [\cite{liu2012emoticon, wang2015sentiment, 10.1145/2339530.2339772, 10.1145/2684200.2684283} ].
The main technique that is used in sentiment analysis in Twitter is to incorporate a lexicon, specially made for the domain of the dataset [\cite{ghiassi2018domain, antonakaki2017social}]. In \cite{kouloumpis2011twitter} they obtain three different corpora of tweets and explore the usage of linguistic features towards sentiment analysis. In \cite{zhang2011combining} they are adopting a lexicon-based method on diverse Twitter datasets.
In \cite{7344886} they conduct a study of sentiment analysis on a dataset of 26,175 general Bulgarian tweets. Through feature selection and classification (binary SVM) they show that the negative sentiment predominated before and after the election period. We are not compiling a specially made lexicon, because we do not have a language barrier or analyzing a plethora of entities; we are just focusing on the two main candidates. 

Sentiment Analysis in Twitter does not have to be limited in one language; there are works studying multiple language datasets [\cite{wehrmann2017character, narr2012language, davies2011language, 7966145, 8122930, saroufim2018language, ptavcek2014sarcasm, 10.1145/3166072.3166082}]. Recent studies in sentiment analysis use Deep Convolutional Neural Networks [\cite{10.1145/2766462.2767830, 8244338, you2015robust, severyn2015twitter, jianqiang2018deep, wehrmann2017character, dos2014deep, alharbi2019twitter, severyn2015unitn}]. Our work focuses on a single language (English) which is the dataset based on. 

There is a plethora of studies that have used sentiment analysis in the political domain [\cite{o2010tweets, antonakaki2016investigating}], either for group polarization [\cite{conover2011political, colleoni2014echo}], during the Arab spring [\cite{weber2013secular}] or for Hugo Chávez [\cite{morales2015measuring}]. For example in \cite{wang2012system} they study a dataset of 36 million tweets on 2012 U.S. presidential candidates and apply a realtime analysis of public sentiment. Using the Amazon Mechanical Turk, they label the dataset with tweets' sentiment (positive, negative, neutral, or unsure) in order to apply statistical classification (Naïve Bayes model on uni-gram features) on a training set consisting of nearly 17.000 tweets (16\% positive,56\% negative, 18\% neutral, 10\% unsure). Also, in \cite{doi:10.1080/1369118X.2013.783609} they attempt to understand the broader picture of how Twitter is used by party candidates, understand the content and the level of interaction by followers. More detailed background on Twitter Sentiment Analysis methods can be found on surveys like \cite{bakshi2016opinion, ravi2015survey, serrano2015sentiment, antonakaki2020survey}.

Background work on sentiment analysis on YouTube [\cite{susarla2012social, wang2011topic}], studies the sentiment on user comments [\cite{thelwall2012commenting}, \cite{lindgren2012took}], identifies the trends and demonstrates the influence of real events of user sentiments [\cite{krishna2014polarity}], implements model utilizing audio, visual and textual modalities as sources of information [\cite{poria2016fusing}] and studies the popularity indicators and audience sentiments of videos [\cite{amarasekara2019exploring}]. YouTube analysis has been used to discover irrelevant and misleading metadata [\cite{bajaj2016disinformation}], to identify spam campaigns [\cite{o2012network}], to discover extremists videos and hidden communities [\cite{sureka2010mining}], to propagate preference information of personalised video [\cite{baluja2008video}], to estimate causality between user profiles [\cite{jang2013deep}] and to apply opinion mining [\cite{severyn2014opinion}]. Our analysis on the YouTube corpus, is based on several metadata from YouTube videos, like user comments, likes and identify the sentiment towards our two main entities: ’Biden’ and ’Trump’.  

\section{Dataset}
\subsection{Twitter}

In this study, we search for all the prevalent hashtags (\#2020election, \#Vote, \#2020usaelection, \#Biden, \#BlueWave2020, \#donaldtrump, \#Election2020, \#Elections\_2020, \#MyPresident, \#November3rd etc.) regarding the US elections on 3 November 2020 and obtain the Twitter corpus
through Twitter API. The acquisition of the dataset started on 19 July 2020 and finished 22 September 2020. This resulted in a dataset of approximately 17.5M tweets, containing 3M users, with the most prevalent hashtags being \#Biden,
\#Trump2020  and \#vote. In table \ref{table:hashtags} we can see the most popular hashtags, sorted by the number of tweets within which
they are contained and in Appendix \ref{appendix}, on table \ref{table:allhashtags} the whole list of hashtags used in our analysis.

\begin{figure}[H]
\centering
\includegraphics[width=\linewidth]{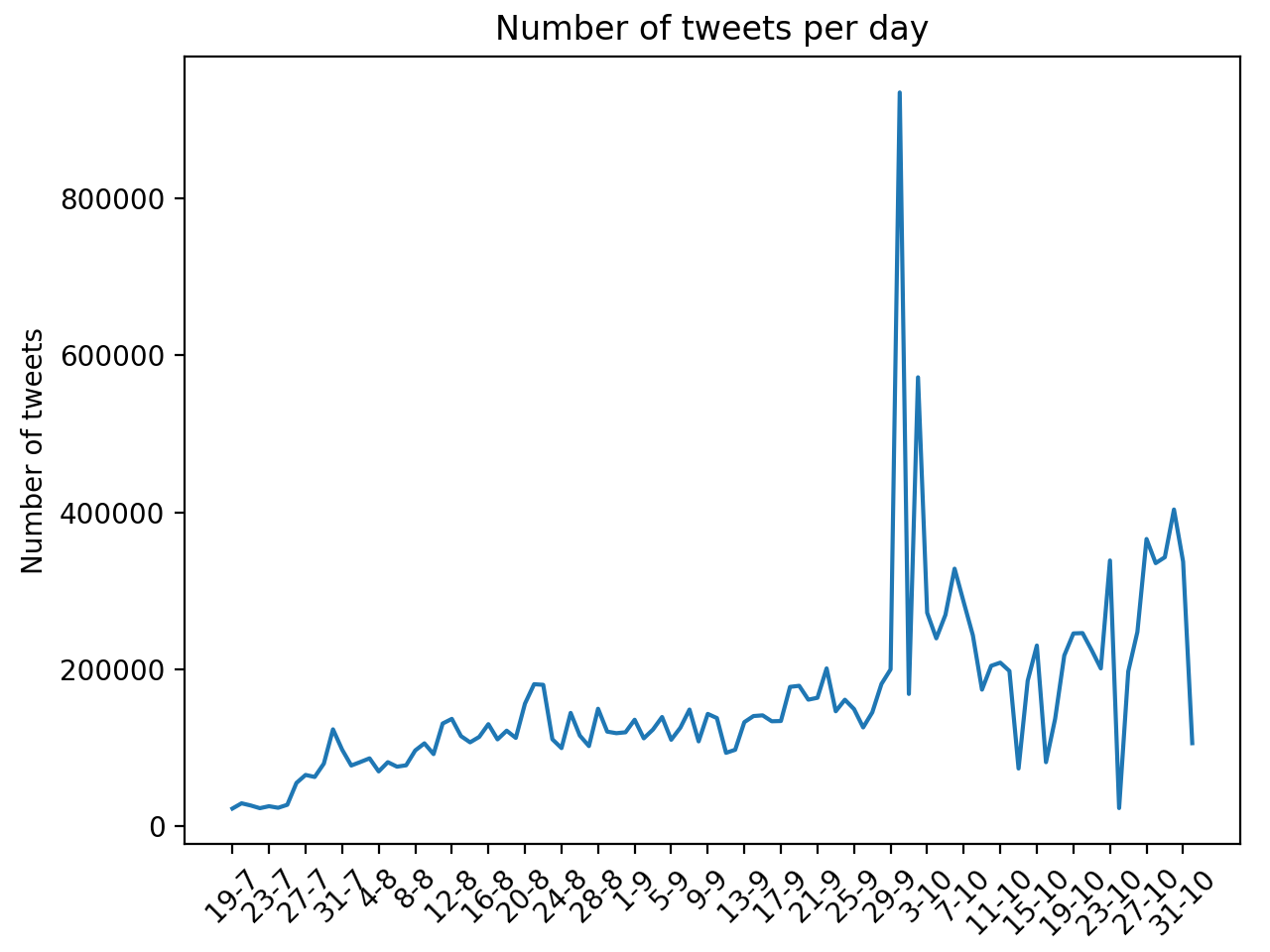}
\caption{Number of tweets per day.}
\label{fig:tweets_per_day}
\end{figure}

\begin{table}[]
\begin{tabular}{|c|c|}
\hline
Hashtag & Tweets count \\
\hline
\#Trump2020 & 2.930.633\\
\#VOTE & 2.258.115\\
\#Vote & 1.785.378\\
\#vote & 1.494.138\\
\#Election2020 &  1.277.839\\
\#Biden & 1.129.063\\
\#Debate2020 & 827.446\\
\#VoteBlueToSaveAmerica & 802.793\\
\#BidenHarris2020 & 658.816\\
\#Trump & 597.172\\
\hline
\end{tabular}
\caption{The 10 most popular hashtags in our dataset.}
\label{table:hashtags}
\end{table}

Figure \ref{fig:tweets_per_day} shows the total number of tweets per day, for every hashtag contained in our dataset.

\subsection{YouTube}
\label{YouTubeMethodology}


From the election tweets, we extracted all the YouTube video links,
contained on those tweets. We followed these links and ended up with 16.642 unique
videos. Through the YouTube Data API, we obtained all the publicly available data
regarding these videos. The accessible data used in this study for each video are:
\begin{enumerate}[label=(\alph*)]
  \item the number of views, likes, dislikes and comments,
  \item the category where it belongs (e.g. News \& Politics, Entertainment, Music, etc.),
  \item the text and the author of each selected comment, and
  \item the YouTube channel that posted the video.
\end{enumerate}

\begin{figure}[!htbp]
\centering
\includegraphics[width=\linewidth]{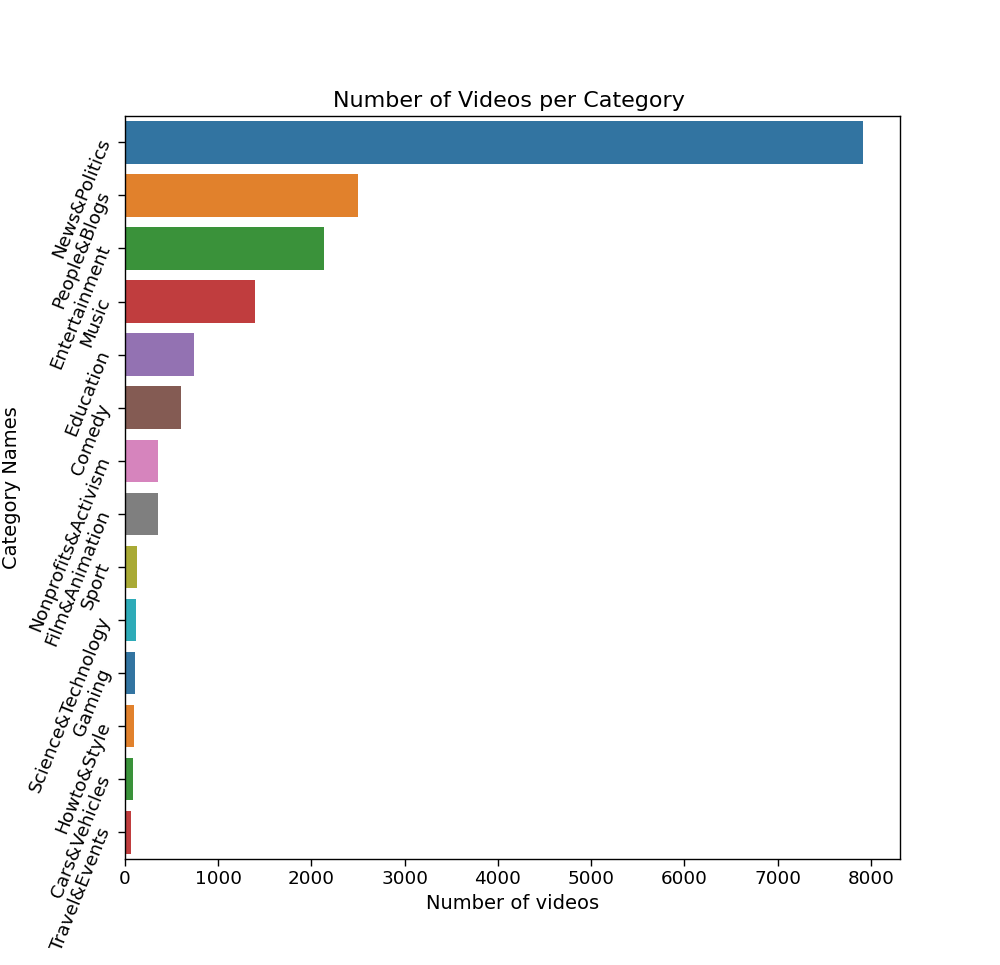}
\caption{Number of videos per YouTube Category}
\label{fig:videoCount}
\end{figure}

In figure \ref{fig:videoCount} we see how many videos belong to each category. In this study we focus on the
election's topic, so we filtered out the videos that do not belong to the following categories:
News \& Politics, People \& Blogs, and Entertainment. The filtering led to a dataset of 12.538 videos.

\begin{figure}[!htbp]
  \centering
  \includegraphics[width=\linewidth]{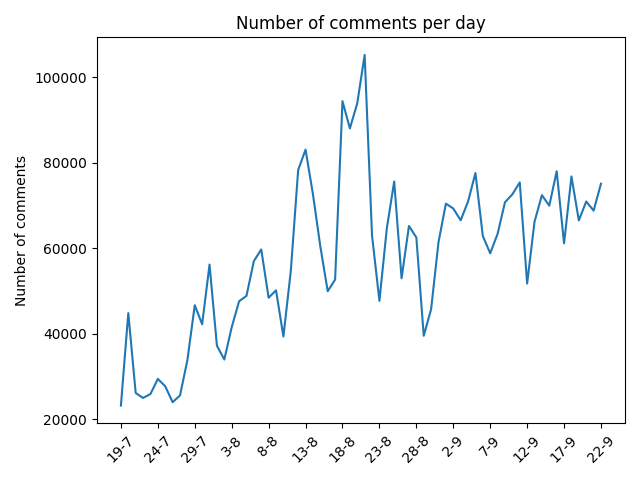}
  \caption{Number of comments in videos per day}
  \label{fig:youtComments}
\end{figure}

From the 12.538 videos, we gathered all the comments and their replies generated between
19/7/20 and 22/9/20. This resulted in a dataset of 3.091.176 unique commenters and 27.927.909
comments and replies. Figure \ref{fig:youtComments} shows the total number of comments
and replies per day, related to the elections. We notice an increase in number of comments from July to September, with diurnal patters.
Also there is a peak in 21/8 potentially explained by a speech from Joe Biden \cite{biden21_aug} as well as three
specific tweets current President Donald Trump posted \cite{trtweet20Aug2020_2, tr20Aug2020, tr20Aug2020_2}.

\begin{figure*}
    \centering
   \begin{tabular}{cc}
     \includegraphics[width=65mm]{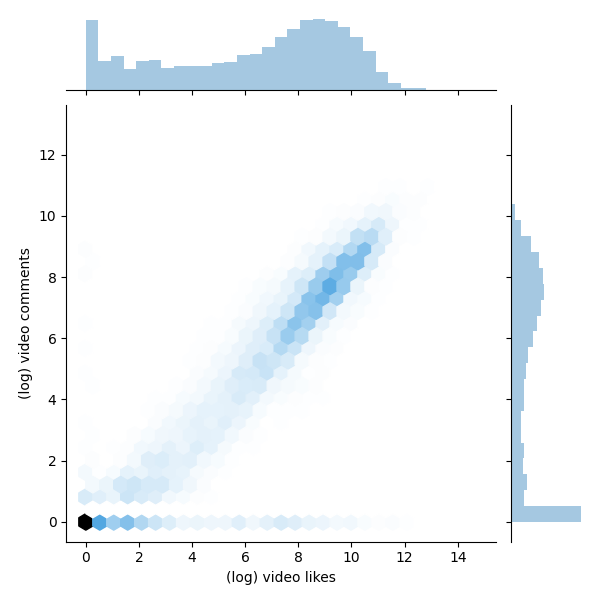} &   \includegraphics[width=65mm]{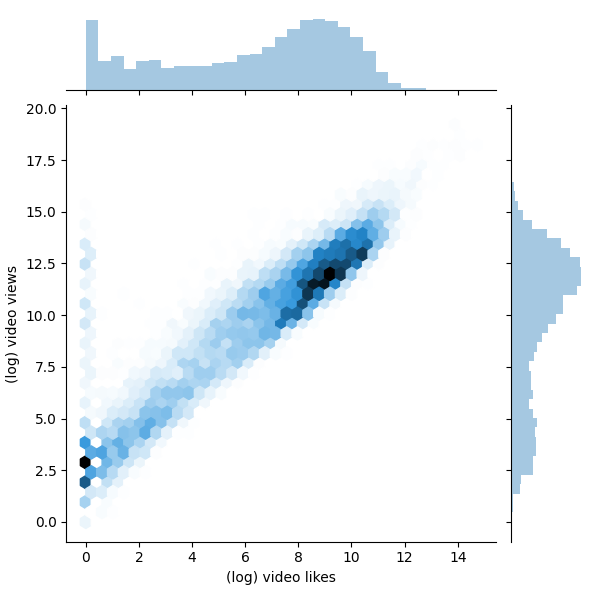} \\
   (a) & (b)  \\[6pt]
     \includegraphics[width=65mm]{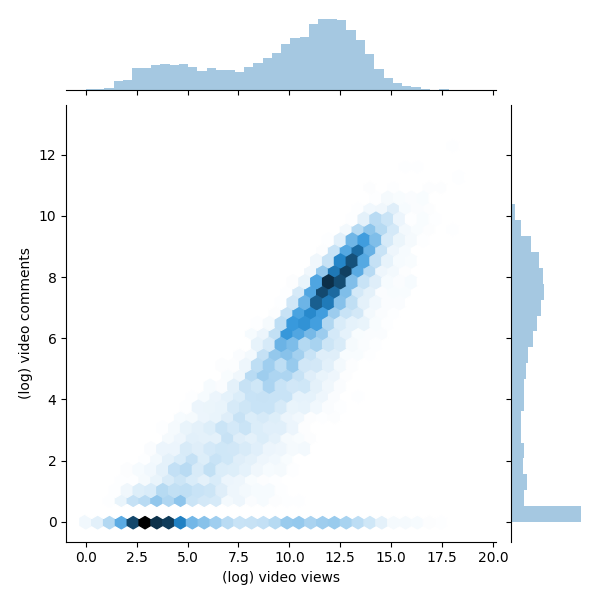} &   \includegraphics[width=65mm]{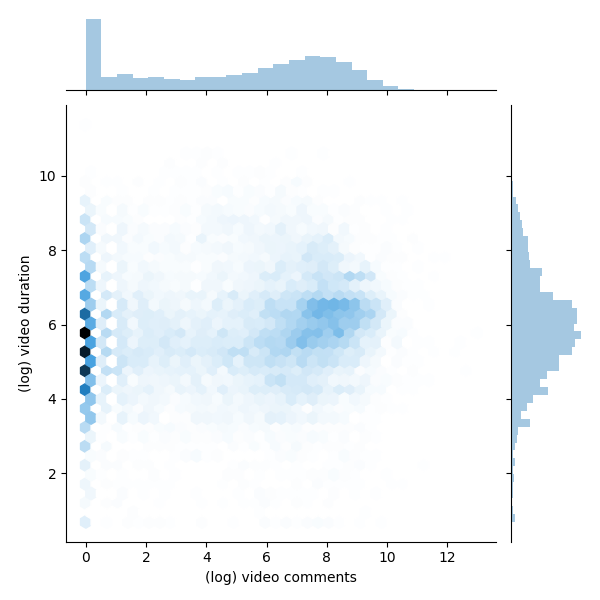} \\
   (c) & (d)  \\[6pt]
    \includegraphics[width=65mm]{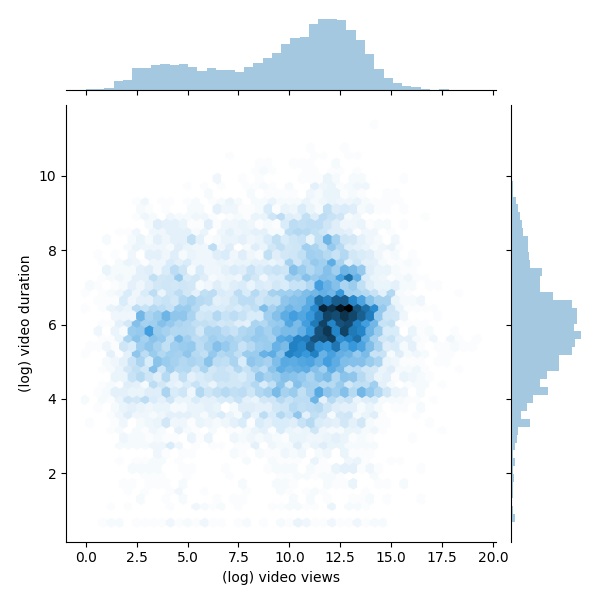} &   \includegraphics[width=65mm]{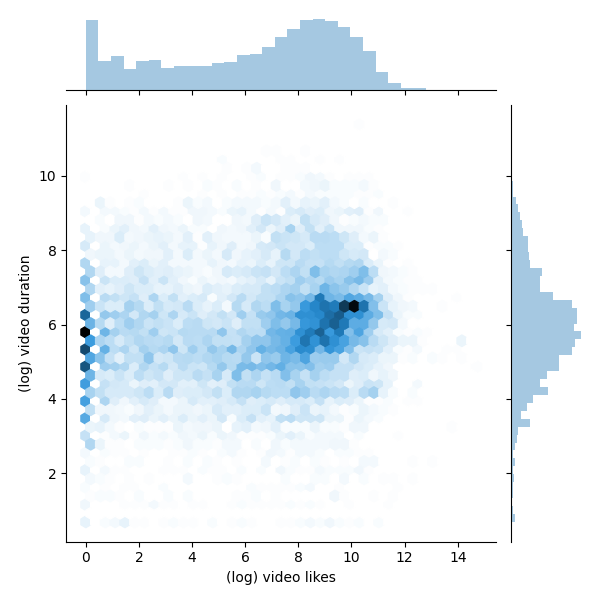} \\
   (e) & (f)  \\[6pt]
   \end{tabular}
   \caption{Join plots that show the relationship between the different features of the videos.}
   \label{fig:youtDataset}
  \end{figure*}

In figure \ref{fig:youtDataset} we see different join plots that illustrate the logarithmic relationship between the accessible features of the 12.538 videos.
For example, in \ref{fig:youtDataset} (a) we see that many videos have 0 comments and close to 0 likes. The main concentration is
between 6 and 10. 
In \ref{fig:youtDataset} (e) we see the comparison of views with the duration of a video. The higher concentration of
views is between 10 and 15 and for the video duration, between 5 and 8.

\section{Methodology}

\subsection{Text Preprocessing}

Initially, for both Twitter and YouTube, we follow a prerequisite set of steps for
preprocessing of the corpus (tweets - comments); by removing punctuation symbols,
text emojis, URLs, by modifying mentions and hashtags and by removing starting
characters of ('@' and '\#'). This procedure removes the text noise and finally
it allows the identification of the entity that was discussed by the users. The next
step includes the transformation of the text to lower-case and the tokenization of
the collected tweets. Then, we perform lemmatization of each token. This technique
normalizes the inflected word forms. As a result of the previous steps, this sentiment
analysis will be performed on lower cased and normalized sentences.

\subsection{Sentiment Analysis}
\label{sentAnalysis}

In our implementation of sentiment analysis, we utilize Vader \cite{gilbert2014vader}
sentiment analysis model from python NLTK library\cite{BirdKleinLoper09}. We choose
this particular implementation because it is especially attuned to the sentiment
expressed in social media. Vader uses a list of lexical features that are labeled
according to their semantic orientation. Thanks to the implementation simplicity,
sentiment analysis execution time remains low. Utilizing the already developed
sentiment solution, we need to perform text filtering and parse our data to
SentimentIntensityAnalyzer function that returns the scores of positive, negative,
and neutral sentiment types. In our analysis, we compute such sentiment for each
collected tweet and comment in our database and summarize those sentiment scores
daily for each particular user. We compute two values of daily sentiment
scores per user: summary, where the daily user sentiment is a sum product of scores
of tweets that was created by the user at each day; and the average score where the
summary score is divided by the number of tweets where a particular entity was mentioned
by the user.

\begin{table}[ht]
\begin{tabular}{|c|c|}
\hline
Entity & Key words\\
\hline
Trump & trump, donald, donaldtrump,\\
      &  trump2020, votetrumpout, \\
      &  trumpislosing, dumptrump, \\
      &  nevertrump, republican\\
\hline
Biden & biden, joseph, bluewave2020,\\
      &  votebluetosaveamerica, voteblue,\\
      &  ridinwithbiden, neverbiden, democrat \\

\hline
\end{tabular}
\caption{The complete list of all entities with the corresponding keywords that were used for each one. }
\label{table:allentities}
\end{table}

\subsection{The Entities}
\label{Entities}
The sentiment analysis of the corpus develops only the emotion vector of a
particular sentence without presenting the entity within it was discussed. To identify
the entity that was described at each tweet, we generate the set of keywords for
the particular dataset entities (`Trump' and `Biden').
We are matching those keywords in each text to identify if the entity was used on the specific
tweet-comment text. Each time the entity is used in a particular tweet-comment, we assign the
sentiment values on those entities to the user who posts this particular tweet.
Those entity sentiments are assigned to the user daily, since we are
interested in the identification of the user/community dynamic and present how entity sentiment
evolves day by day.
The two entities along with the corresponding keywords that were
used for each one, are on table \ref{table:allentities}.

\subsection{Event consequences}
\label{EventConsiquences}
Since our work is based on the analysis of the 2020 US Presidential elections we monitor real world events that may trigger significant user interest in social media. Interesting examples of such events that can be used are the candidates' debate on TV. The depict of the online conversations regarding these events are visible within our analysis. 
By analyzing the user engagement of these specific periods, one day before and one day after, shows whether such real world events are connected in the virtual world of social networks. For this reason we selected
all TV debate dates and the dates after President trump was diagnosed positive with COVID-19.

\section{Results}

\subsection{Volume Analysis}

\begin{figure}[!htbp]
\centering
\includegraphics[width=\linewidth]{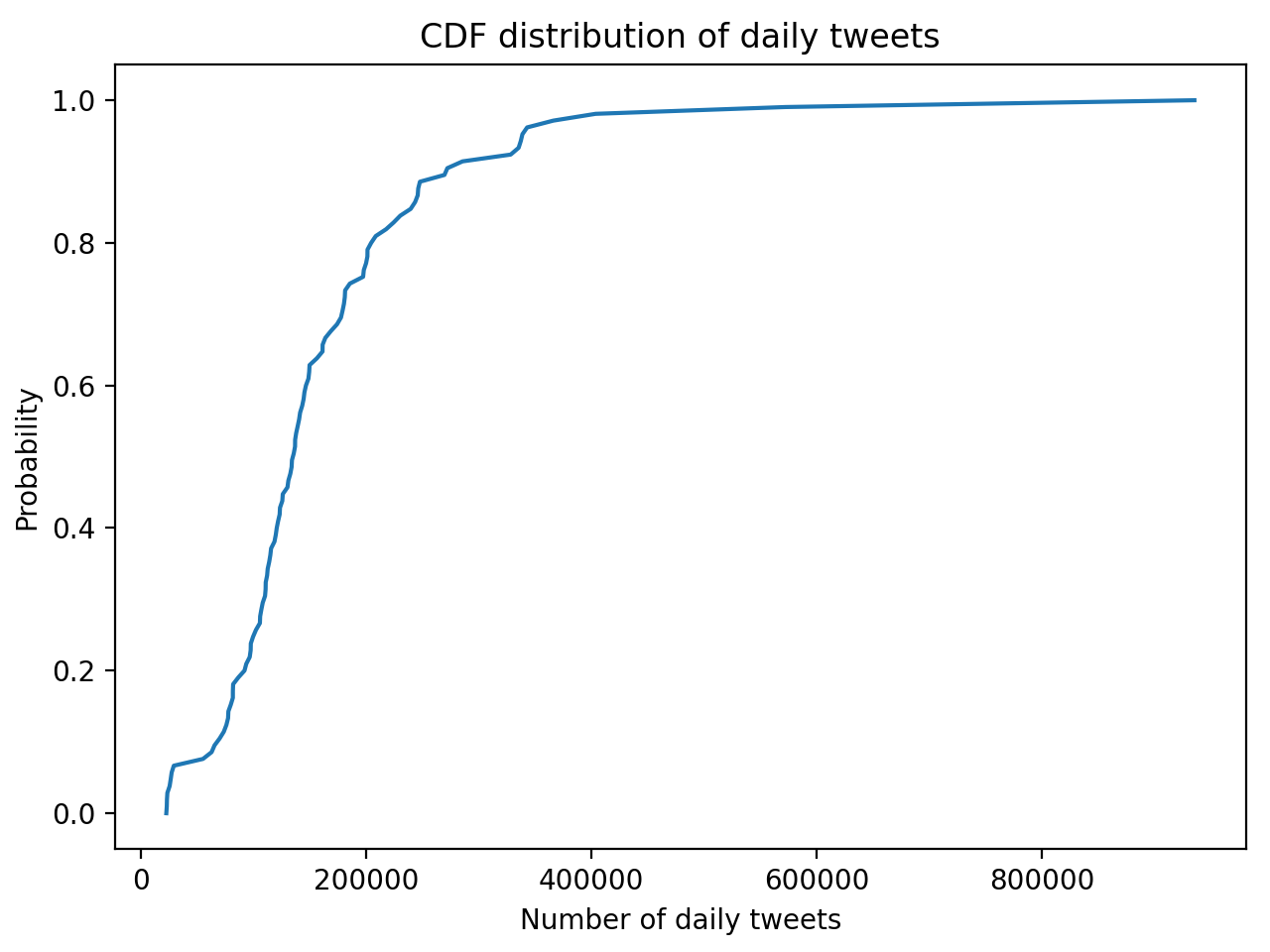}
\caption{CDF of tweets per day}
\label{fig:tweets_per_day_cdf}
\end{figure}

\begin{figure}[H]
\centering
\includegraphics[width=\linewidth]{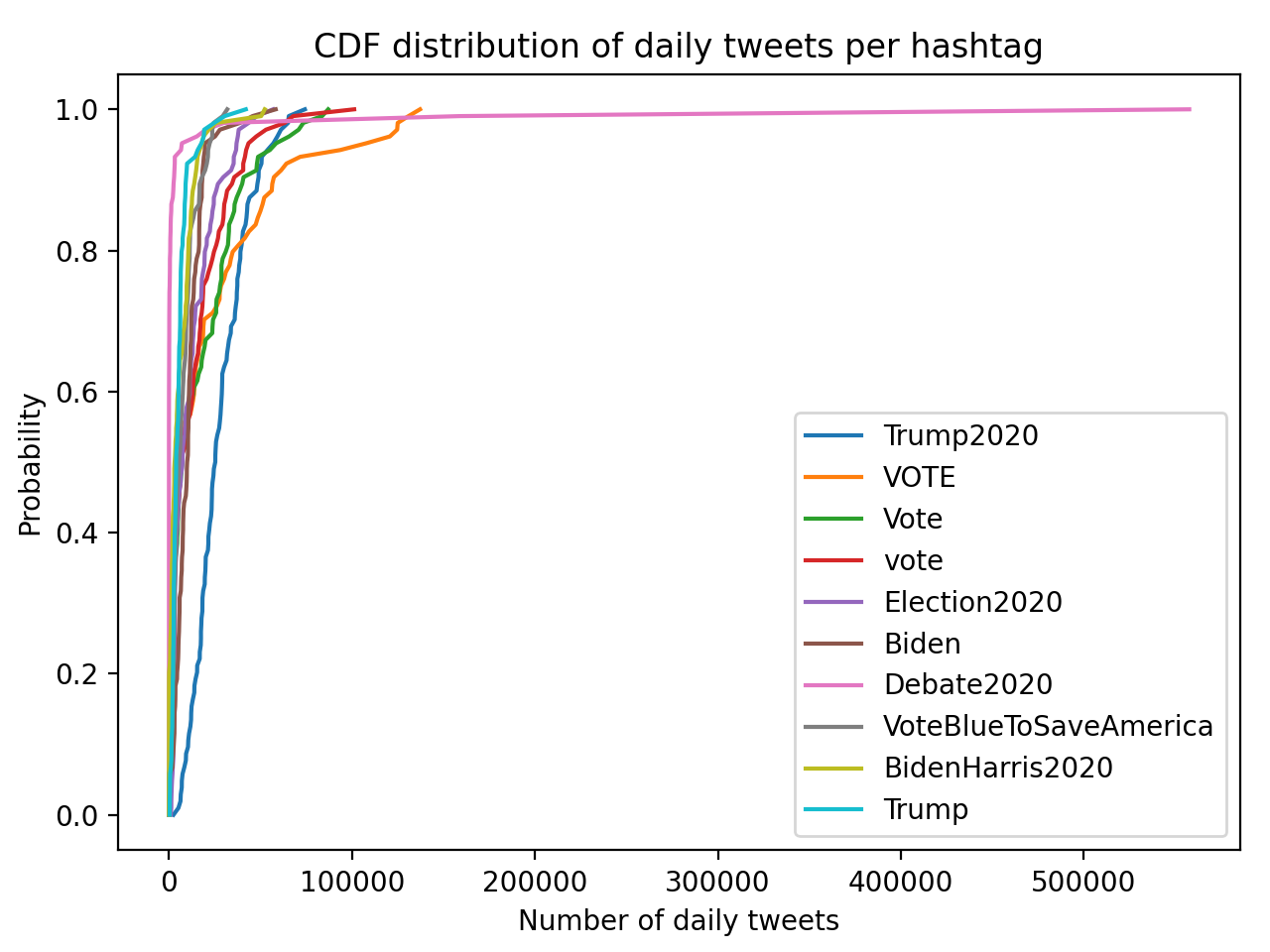}
\caption{CDF distribution of the daily tweets per hashtag.}
\label{tweets_per_hashtag_cdf}
\end{figure}

In this section, we include several volume measures derived from our dataset.
Initially, we perform a volume analysis on the whole corpus of the tweets.
In figure \ref{fig:tweets_per_day_cdf} we plot the cumulative distribution of the
daily tweets where we notice that 90\% of the times there are 150.000 tweets daily post. In figure \ref{fig:tweets_per_day}  we plot the number of tweets per day where we notice the diurnal increases of the posts. In figure \ref{tweets_per_hashtag_cdf}
we can see the cumulative distribution of the daily tweets per hashtag, where we note the
prevailing hashtags of \#DonaldTrump, \#MAGA, \#vote and \#Trump.

\begin{figure}[!htbp]
\centering
\includegraphics[width=\linewidth]{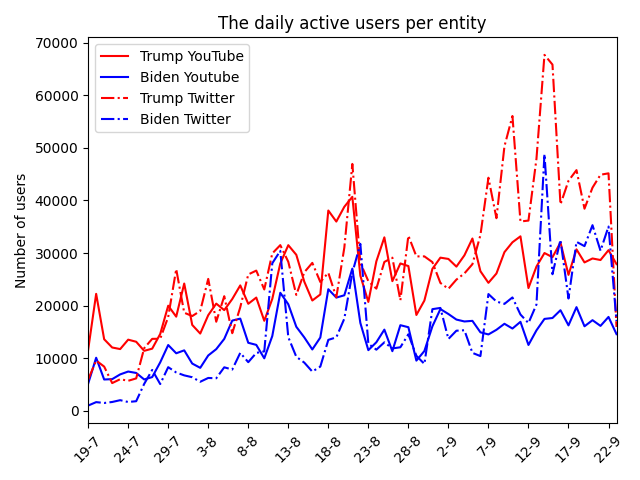}
\caption{In this figure, we represent the number of daily users that post
YouTube comments and tweets where particular entities are discussed.}
\label{fig:daily_users_per_ent_Youtube}
\end{figure}

We plot the daily active users per entity for both Twitter and YouTube in figure \ref{fig:daily_users_per_ent_Youtube},
for each entity of `Trump' and `Biden' and we notice that the total number of users are increasing from July to September and that
the users posting at Twitter and commenting on YouTube for Trump exceed the users for Biden, in Twitter and YouTube respectively.
The number of users posting tweets for the entity of Trump seems to exceed the corresponding number of users for the entity `Biden'.
In both figures, we notice a peak on day 20/8/2020, potentially because of three specific tweets
current President Donald Trump posted \cite{trtweet20Aug2020_2, tr20Aug2020, tr20Aug2020_2}. 
We also notice the total traffic of Twitter to overcomes the total YouTube traffic.

\subsection{The retweet graph}

\begin{figure*}
 \centering
\begin{tabular}{cc}
  \includegraphics[width=65mm]{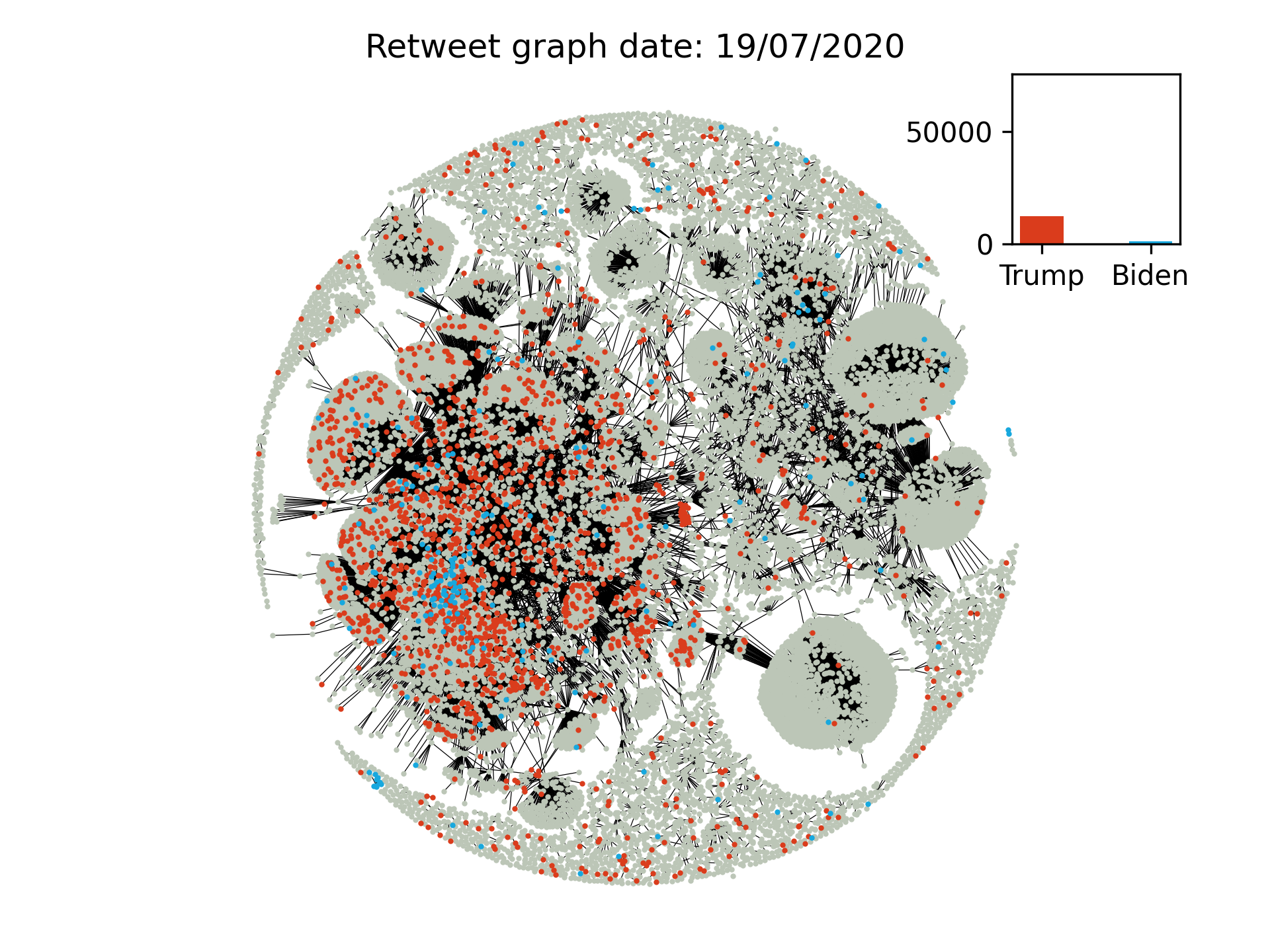} &   \includegraphics[width=65mm]{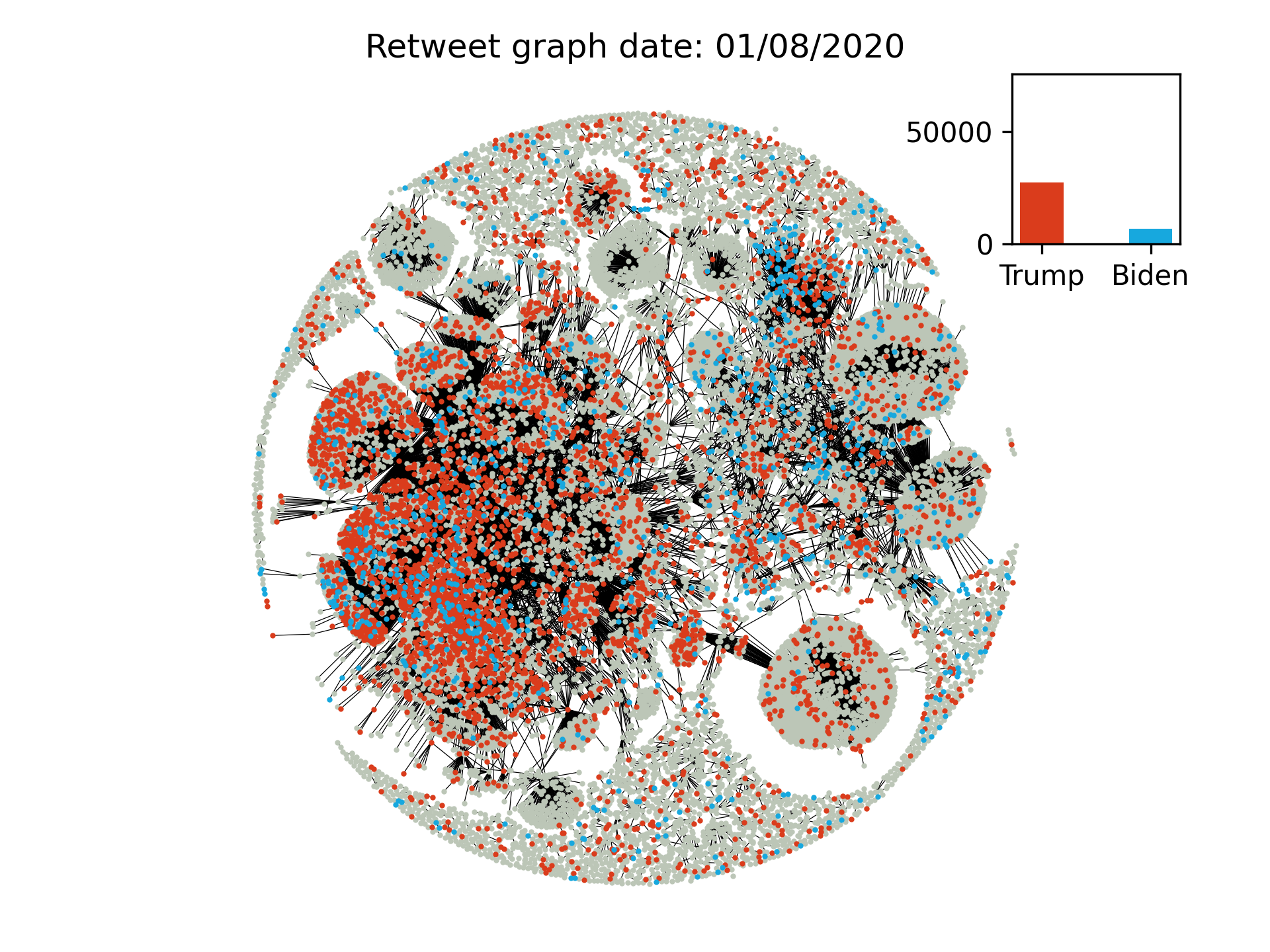} \\
(a) & (b)  \\[6pt]
  \includegraphics[width=65mm]{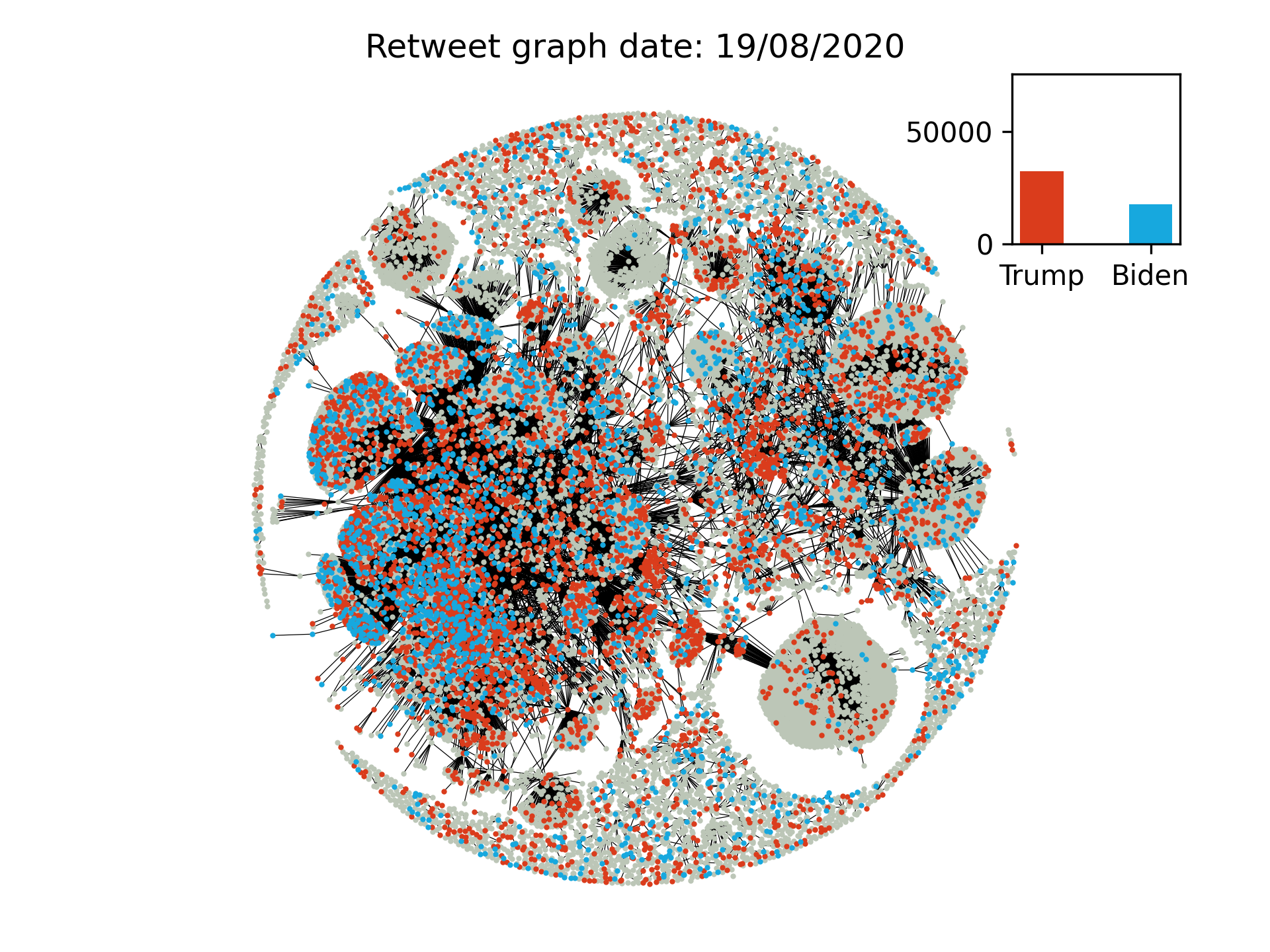} &   \includegraphics[width=65mm]{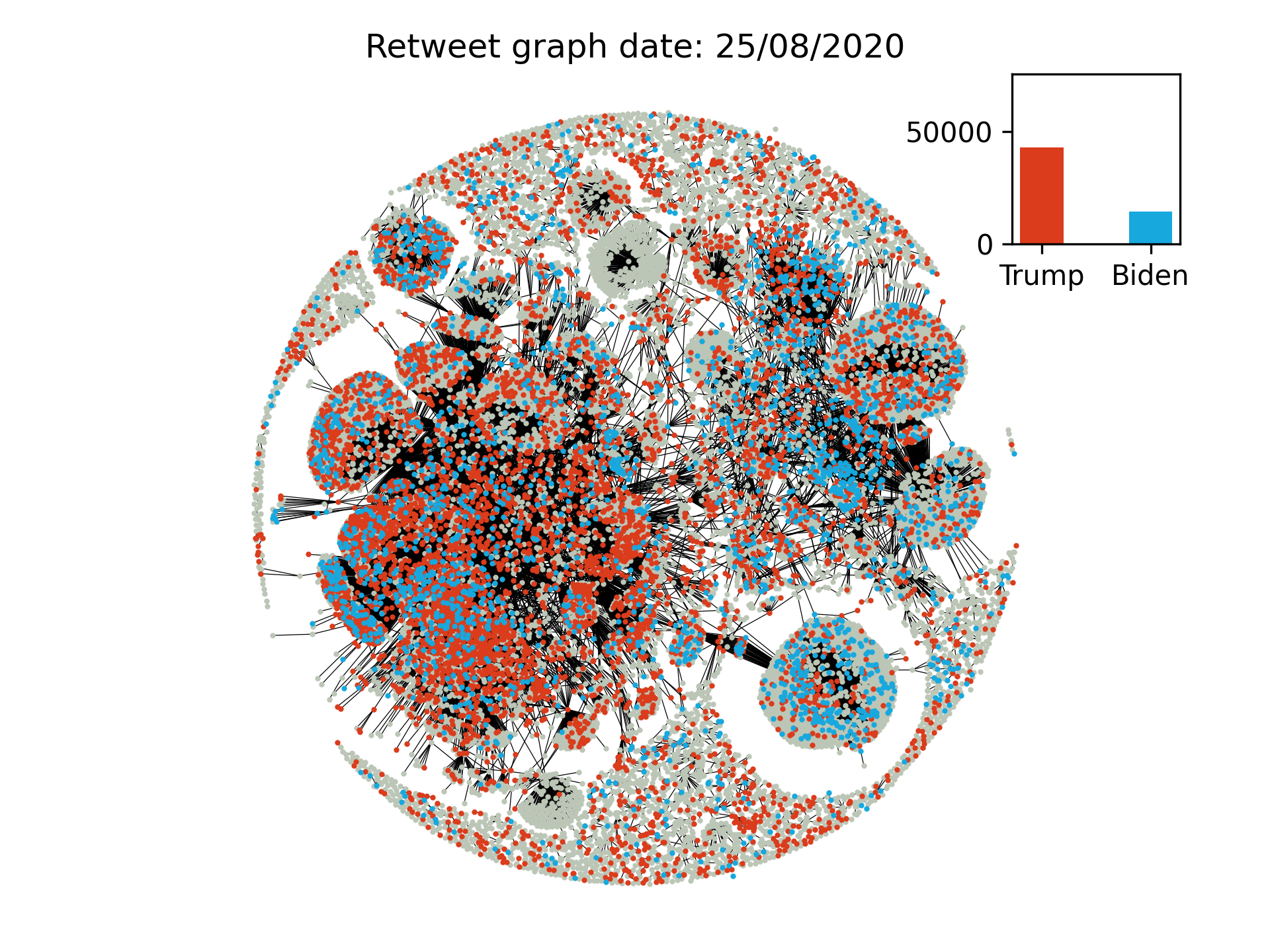} \\
(c) & (d)  \\[6pt]
 \includegraphics[width=65mm]{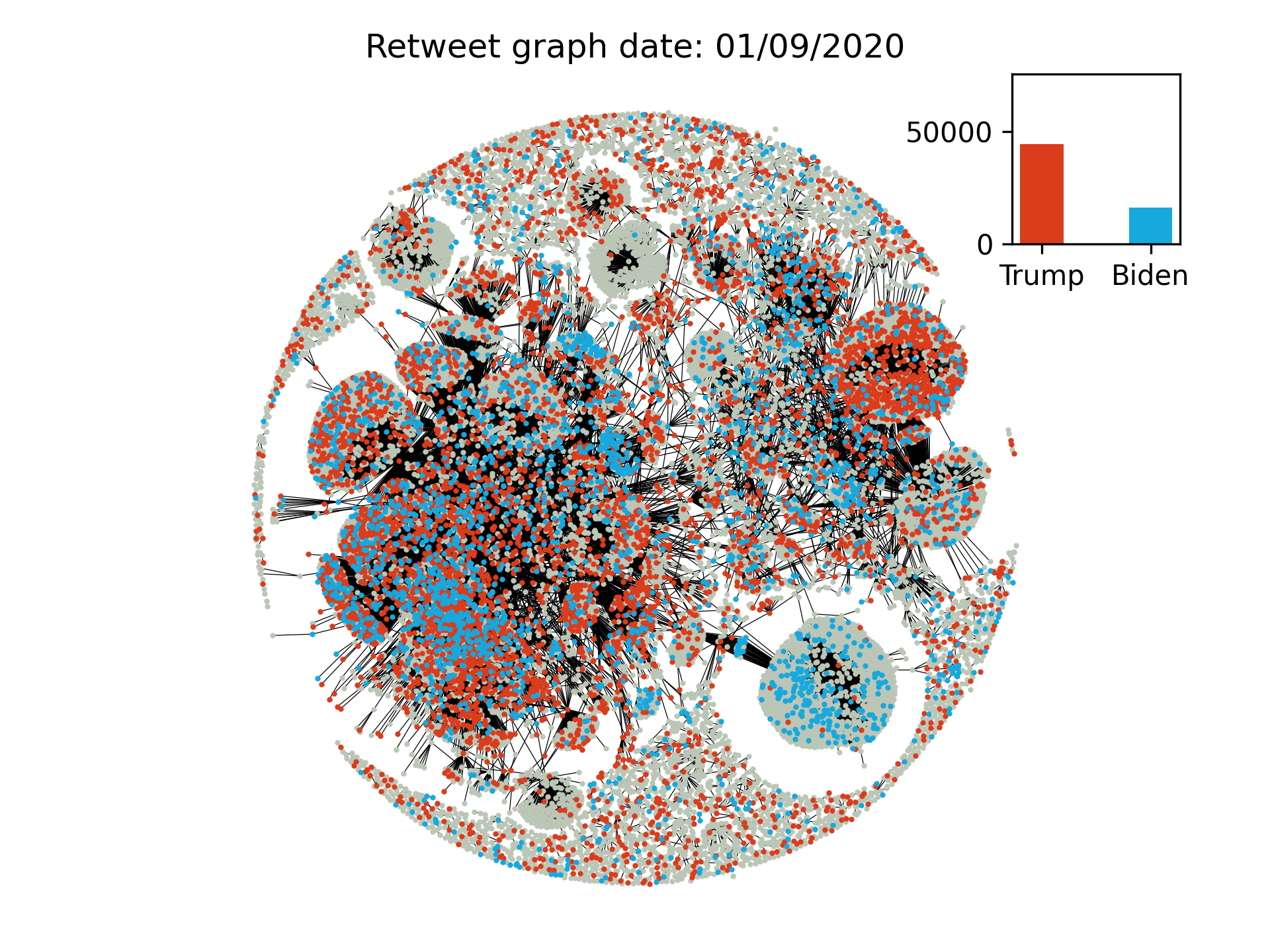} &   \includegraphics[width=65mm]{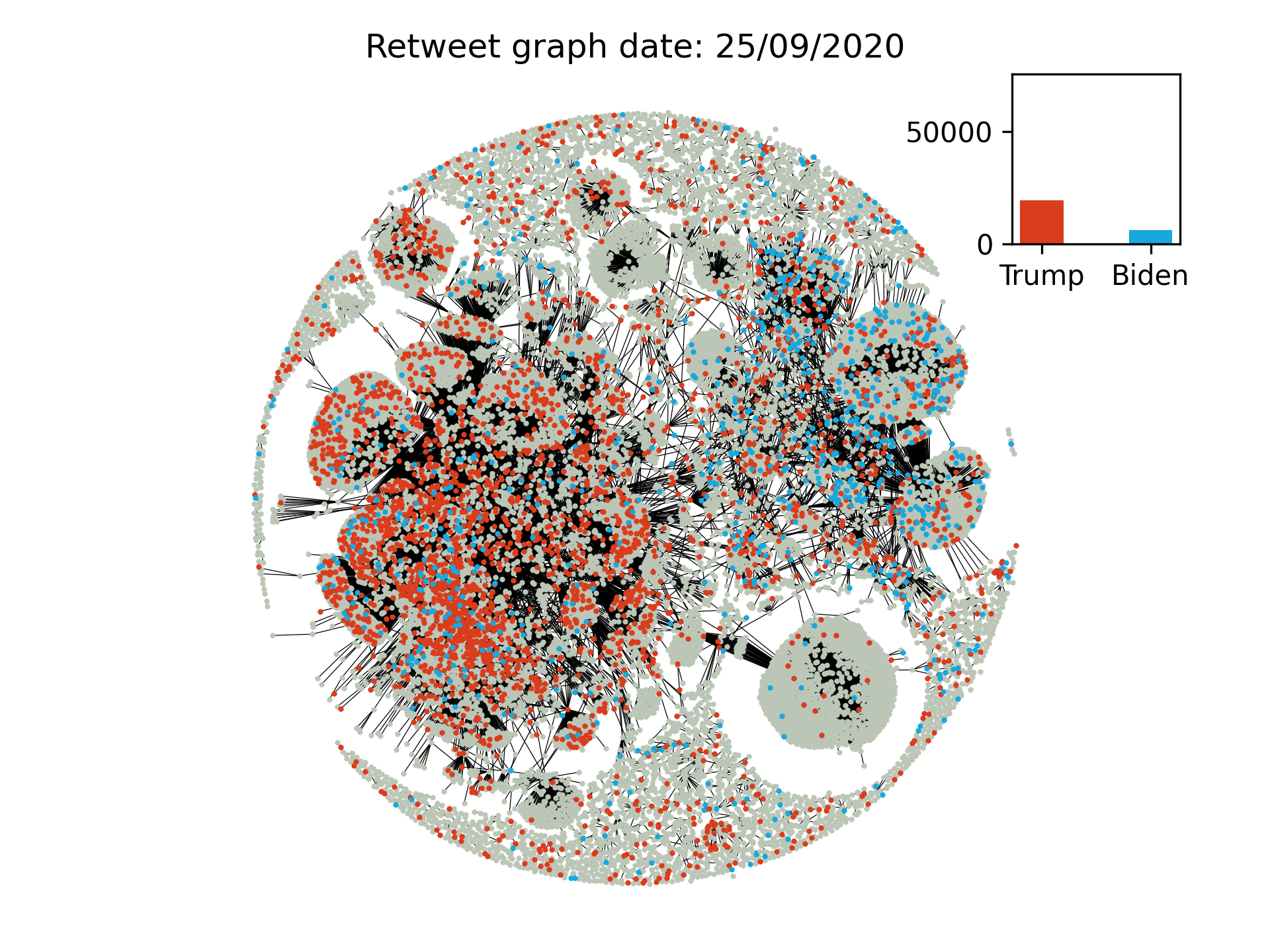} \\
(e) & (f)  \\[6pt]
\end{tabular}
\caption{Retweet graph relation between users in our dataset. Colours represent
entity activity by user red colour represent entity Trump and blue colour entity Biden.}
\label{retweet_graph}
\end{figure*}

In figure \ref{retweet_graph} we plot the retweet graph.
This graph is developed by the retweet relationships between all users in the collected dataset.
We present the retweet relation as a directed edge between two users (nodes). Also, we assign
edge weight with the number of retweets that users are performed for a particular
destination node.  We also perform a filtering metric where we remove the low
weighted edges of weight 1 until weight 7. With the filtering procedure, we reduce the noise
the volume of not significant relations and also reduce the number of edges that are manageable for visualization.
Before filtering, our graph consisted of 1.142.376 nodes and 3.859.640 edges and after filtering non-significant edges
we reduce the number of nodes to 114.971 and the number of edges to 211.534.
For the entity visualization, we use 2 colors for our entities, in red color
represent the entity of Trump, and in blue color the entity of Biden.

We apply sentiment analysis on each day and
we measure the number of tweets that a user posts containing a particular entity. We use this
counter to provide coloring of the user node by selecting the most popular
entity of the user at each particular date. We also develop a bar plot with the volume of users
that allows us to compare the daily volume per entity.
Users that don't use any entities in their tweets, remain without particular coloring.

Graph plots \ref{retweet_graph} were generated with Gephi Furchterman Reingold layout \cite{ICWSM09154}
while we export Gephi generated positions and we use them to generate a daily graph with networkX python
library \cite{networkx_lib}. In tables \ref{table:indegreeHigh} and \ref{table:outdegreeHigh} we show
the highest retweeted used with the highest numbers of in degree and out-degree respectively, with anonymized usernames.

\begin{table}[ht]
\begin{tabular}{|c|c|c|}
\hline
User  & Number of Followings & Number of Followers  \\
\hline
User 1 & 62.9K  & 68.5K \\ 
User 2 & 18.5K  & 78.4K \\ 
User 3 &  43.1K & 61.7K \\ 
\hline
\end{tabular}
\caption{Top retweeted users(highest indegree).}
\label{table:indegreeHigh}
\end{table}

\begin{table}[ht]
\begin{tabular}{|c|c|c|}
\hline
User  & Number of Followings & Number of Followers  \\
\hline
User 1 & 5,299 & 8,505 \\ 
User 2 & 37.1K & 34.7K  \\ 
User 3 &  72.3K & 78.1K \\ 
\hline
\end{tabular}
\caption{Top retweeted users(highest out degree).}
\label{table:outdegreeHigh}
\end{table}



\subsection{Sentiment Analysis}
\label{resultssentAnalysis}

\begin{figure}[!htbp]
\centering
\includegraphics[width=\linewidth]{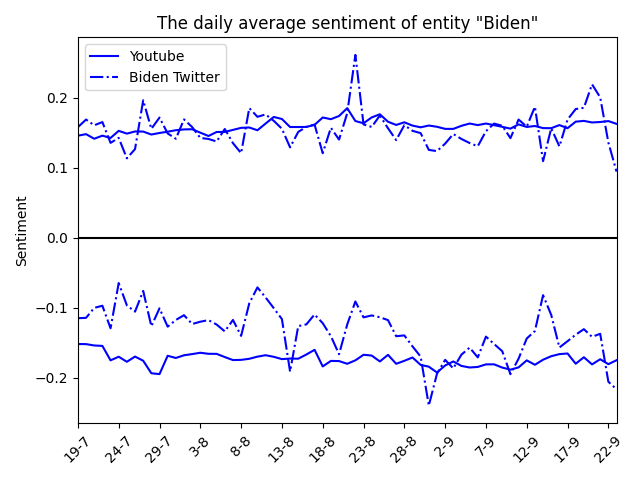}
\caption{Result of daily average sentiment for the entity for Twitter and YouTube collected dataset
for entity Biden.}
\label{sent_per_day_biden_avg}
\end{figure}

\begin{figure}[!htbp]
\centering
\includegraphics[width=\linewidth]{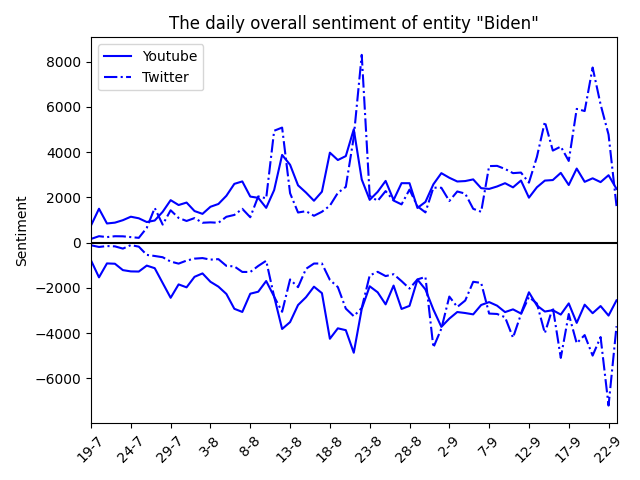}
\caption{Result of overall sentiment analysis for Twitter and YouTube collected dataset
for entity Biden. Particular sentiment values present overall sentiment that
was posted by users at given dates.}
\label{sent_per_day_biden_sum}
\end{figure}

In this section, we present the results from the sentiment analysis in our corpus, as described in \ref{sentAnalysis}.
In figure \ref{sent_per_day_biden_avg} we present the daily average sentiment for the entity `Biden'.
The solid line is the average sentiment in YouTube comments and the dotted line
is the average sentiment in the corpus of the tweets. Below zero we have the
negative sentiment and above zero the positive sentiment for each social media.
On 23 August of 2020, we notice a peak on the positive sentiment and could be
potentially explained by \cite{biden23_8, biden23_8_2}.
In figure \ref{sent_per_day_biden_sum}, we plot the daily overall sentiment for entity Biden,
where we notice the same peak of negative sentiment on 23 August, while a second peak on 19/9 until 22/9
could be explained by his tweet regarding the successor of Justice Ginsburg \cite{biden_20_9} or a follow tweet on the current president\cite{biden_20_9_2}.

\begin{figure}[!htbp]
\centering
\includegraphics[width=\linewidth]{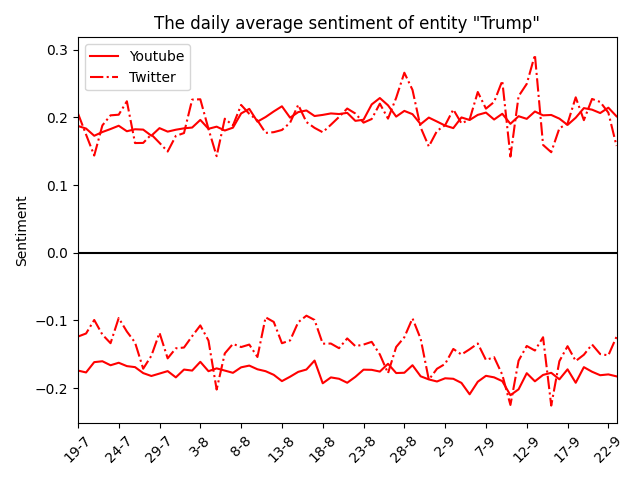}
\caption{Result of daily average sentiment for the entity for Twitter and YouTube collected dataset
for entity Trump.}
\label{sent_per_day_trump_avg}
\end{figure}

\begin{figure}[!htbp]
\centering
\includegraphics[width=\linewidth]{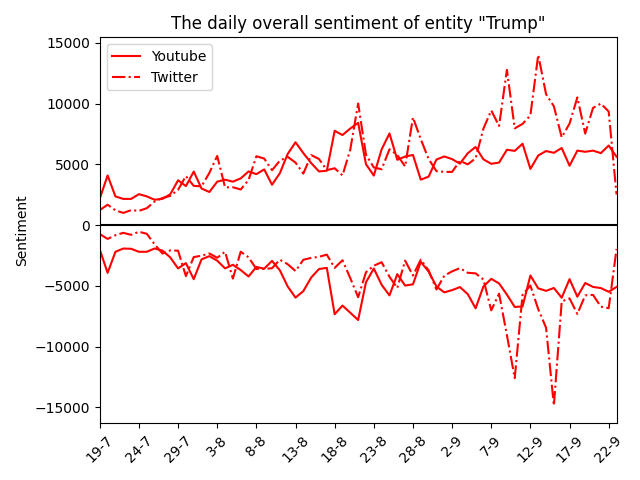}
\caption{Result of overall sentiment analysis for Twitter and YouTube collected dataset
for entity Trump. Particular sentiment values present overall sentiment that
was posted by users at given dates.}
\label{sent_per_day_trump_sum}
\end{figure}

The corresponding daily plots for overall and average sentiment on the entity of Trump are shown
in \ref{sent_per_day_trump_sum} and \ref{sent_per_day_trump_avg}. The peak on overall positive sentiment on
12-13 September could be explained by \cite{trump_1, trump_3, trump_4, trump_5}.


\begin{figure}[!htbp]
\centering
\includegraphics[width=\linewidth]{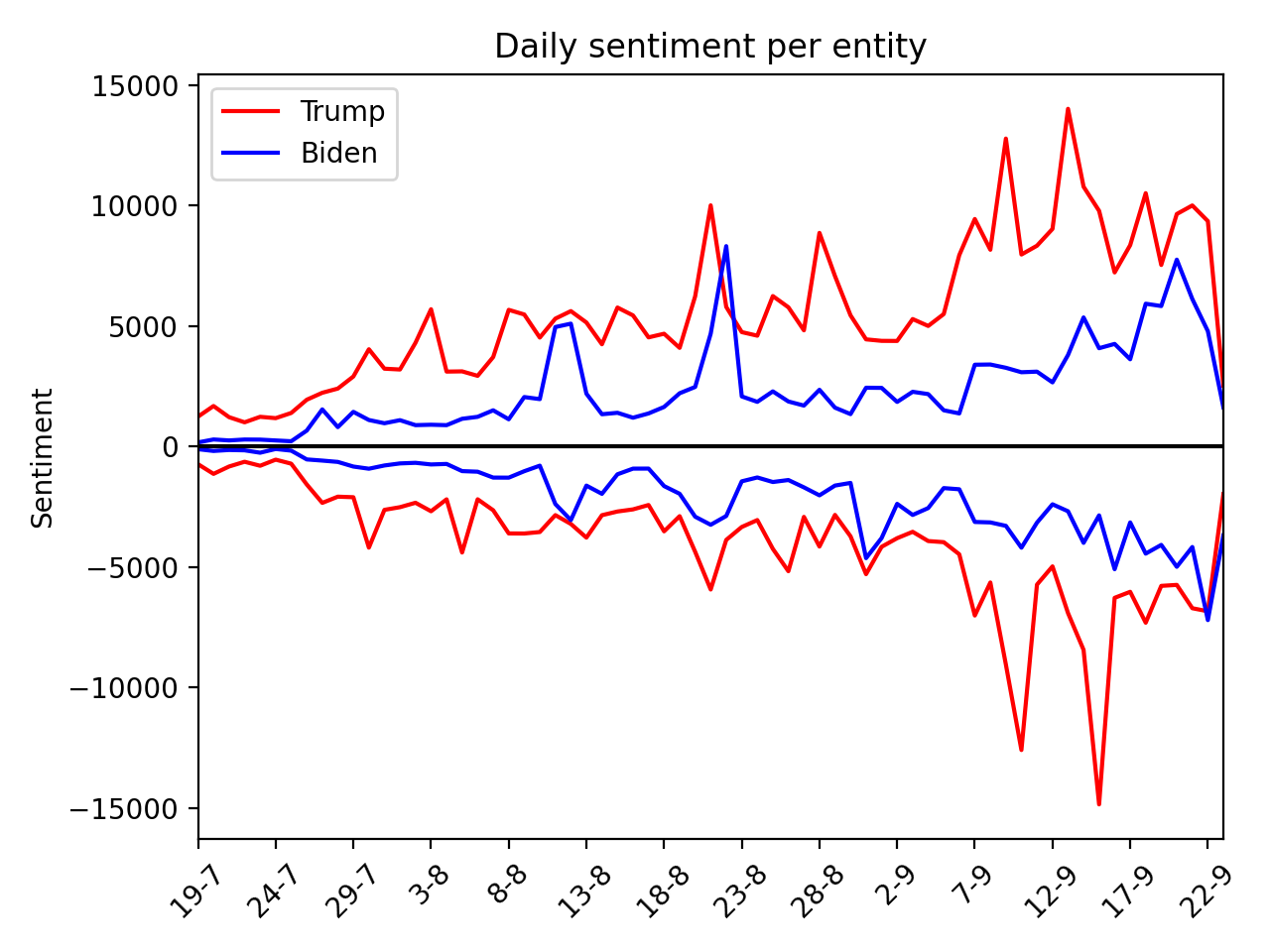}
\caption{Result of daily sentiment per entity.}
\label{sentiment_per_entity_line}
\end{figure}

\begin{figure}[!htbp]
\centering
\includegraphics[width=\linewidth]{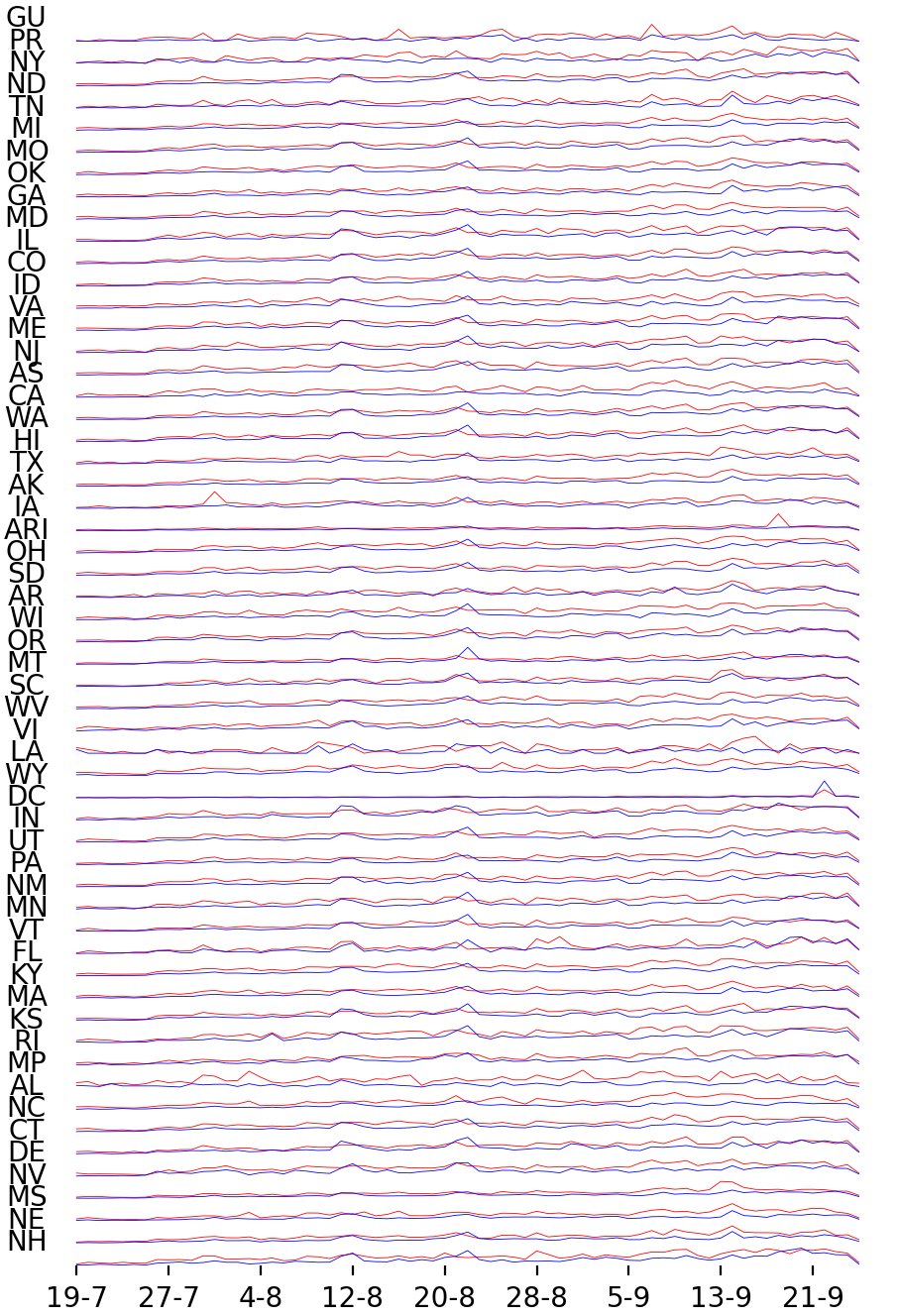}
\caption{Positive sentiment time series for the two sets of hashtags for each state}
\label{sentiment_timeseries}
\end{figure}

Additionally, in figure \ref{sentiment_timeseries} we plot the positive sentiment time series for the two sets of hashtags for each state. We notice the daily fluctuations for every states per entity (blue is the entity ’Biden’ and red is for ’Trump’. The juxtaposition of the time series in the form resembling an EEG makes it easier to discern localized events from nation-wide twitter traffic.
The list of states abbreviations can be found here: \cite{states_abbr}.

s

\begin{figure}[!htbp]
\centering
\includegraphics[width=90mm,scale=0.5]{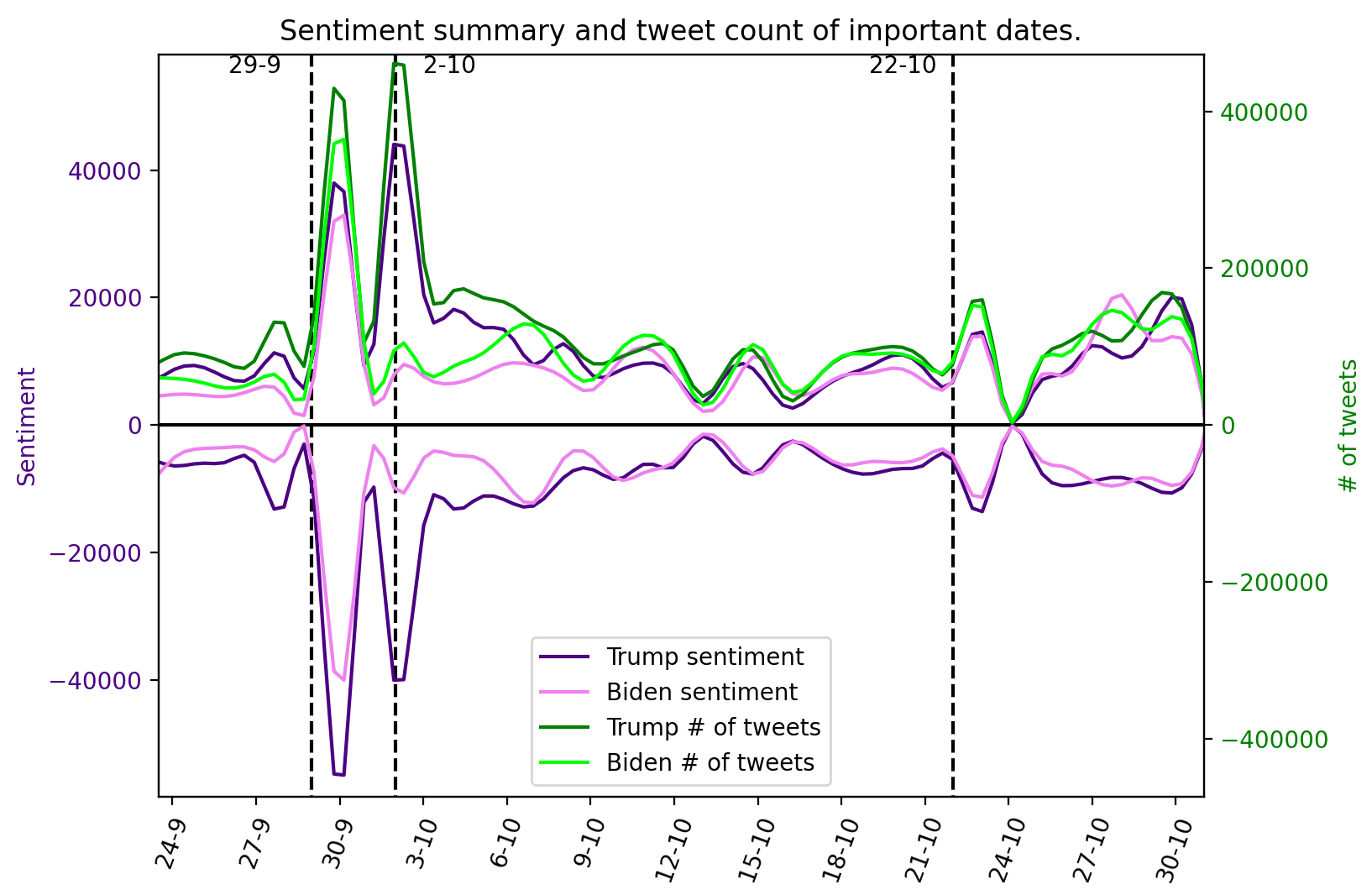}
\caption{Summary of user tweets sentiment values per each entity followed by
number of tweet. At presented plot we mention 3 important dates in dataset, where
29-09 is the date of first debate, 2-10 the date when Trump was tested positive
to COVID-19 and 22-10 the date of last election debate.}
\label{sentiment_sum_count}
\end{figure}

\begin{figure}[!htbp]
\centering
\includegraphics[width=90mm,scale=0.5]{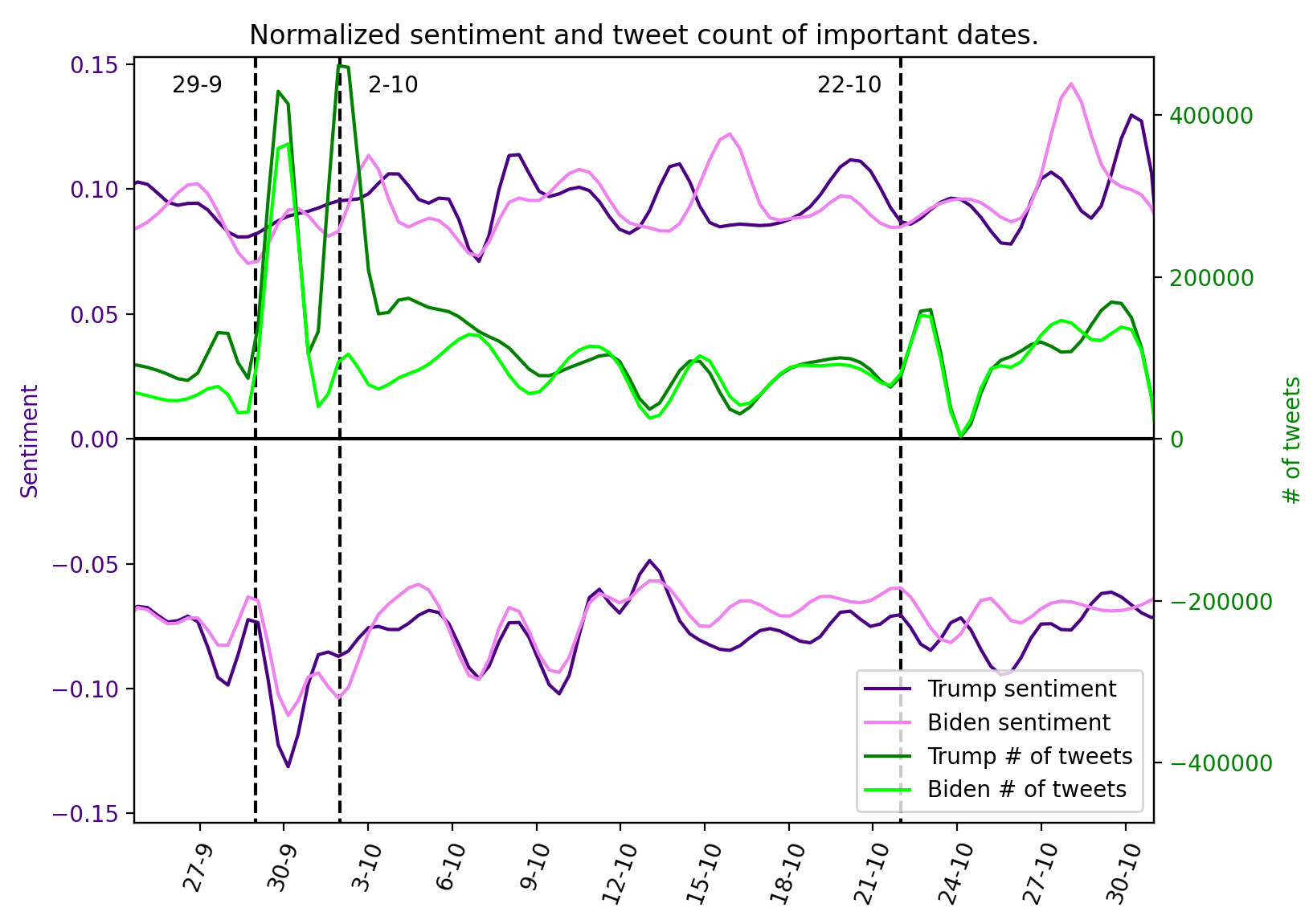}
\caption{Normalized sentiment values of user tweet per each entity followed by
number of tweet. At presented plot we mention 3 important dates in dataset, where
29-09 is the date of first debate, 2-10 the date when Trump was tested positive
to COVID-19 and 22-10 the date of last election debate.}
\label{sentiment_norm_count}
\end{figure}

\subsection{Event consequences}
\label{resultsEventConsiquences}

We use sentiment analysis on specific time points of our dataset timeline ,with important events that allow us to identify how the social media users
react to those events. Specifically, we select the dates of the TV debates
(September 29 and October 22) and the date when the President Trump was
diagnosed positive to COVID-19 (October 2). We selected those particular dates since they were the most important events during the pre-elections period. We use the sentiment analysis results to identify the fluctuations of the sentiment and we measure the
volume around each entity topic.
Our results are presented in figures: [\ref{sentiment_norm_count},\ref{sentiment_sum_count}], where it is
noticeable that the first debate and Trump COVID-19 announcement events generated
high volume of user interest in social media. In the first case of the debates both
entities are soared in comparison with previous dates. At the second event, the volume
of the entity `Trump' is taking the first place during the user discussions on social media
by increasing the volume of tweets for the entity `Trump' and by presenting high dissonance on sentiment values.

\section{Linking Twitter and YouTube data}
In this section, we explore the differences in discussion and
community between the social networks.  We perform Louvain community
detection on both social graphs, we associate communities in the
YouTube comment graph with communities in the Twitter retweet graph,
and measure their similarity and differences.

\begin{figure}
\includegraphics[width=90mm]{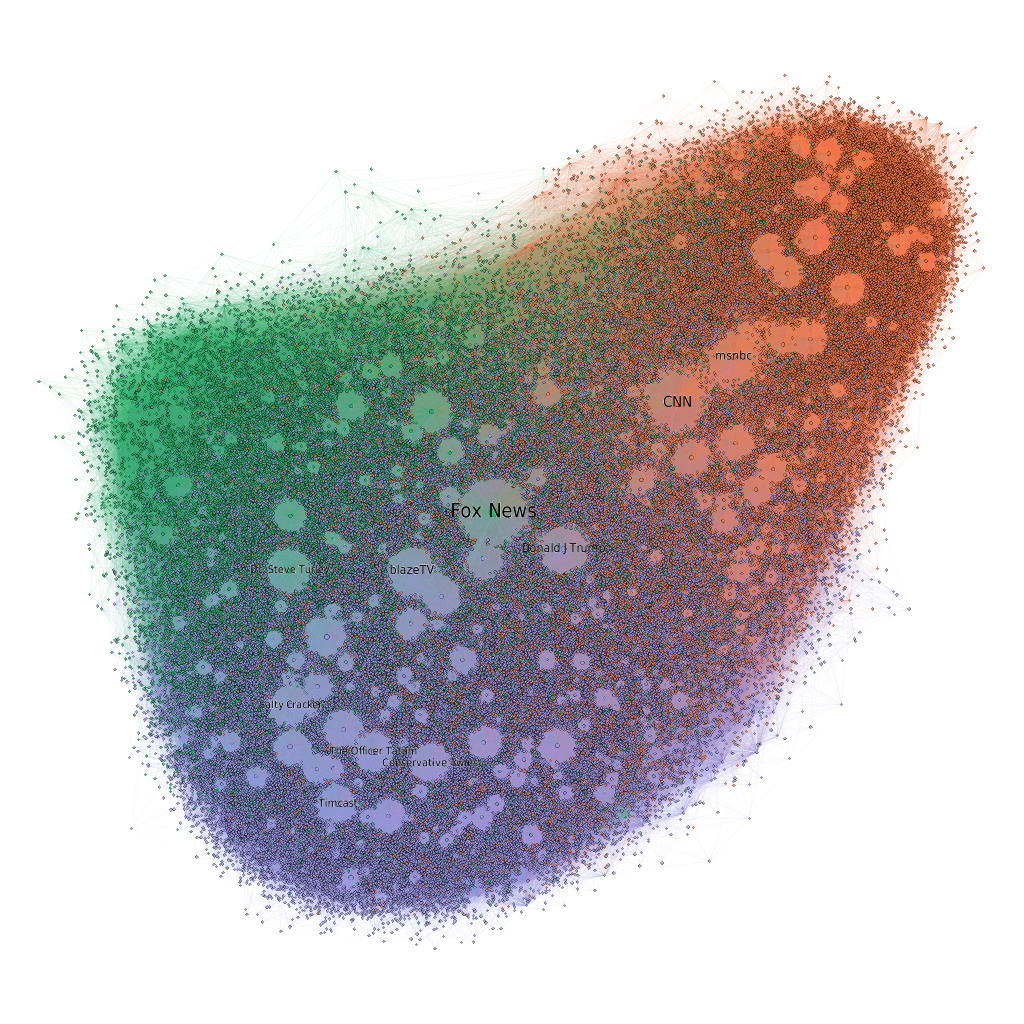}
\caption{The 4-core of the YouTube comment graph, color-coded by community.}
\label{fig:yt}
\end{figure}

Figure~\ref{fig:yt} shows the 4-core of the YouTube comment graph.

\begin{figure}
\includegraphics[width=90mm]{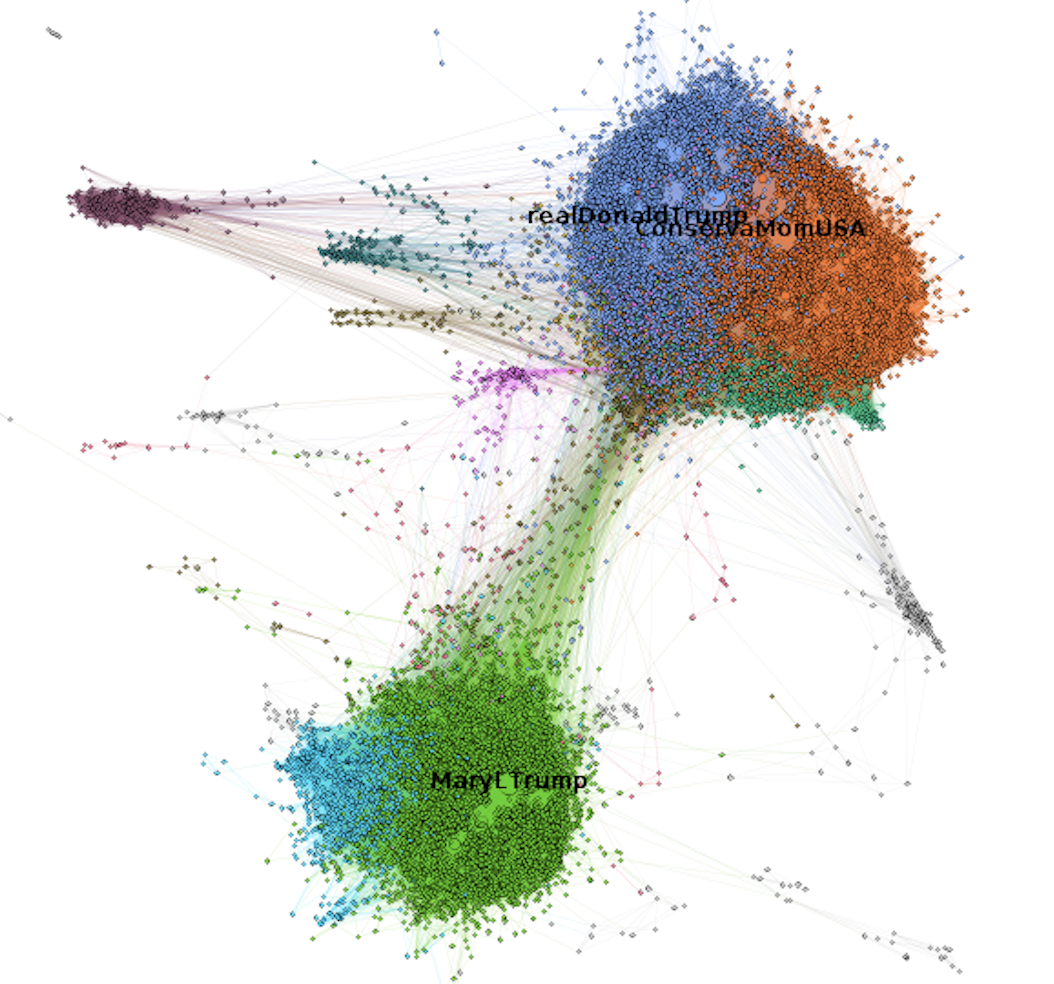}
\caption{The 4-core of the YouTube comment graph, color-coded by community.}
\label{fig:rt}
\end{figure}

\begin{figure}
\includegraphics[width=90mm]{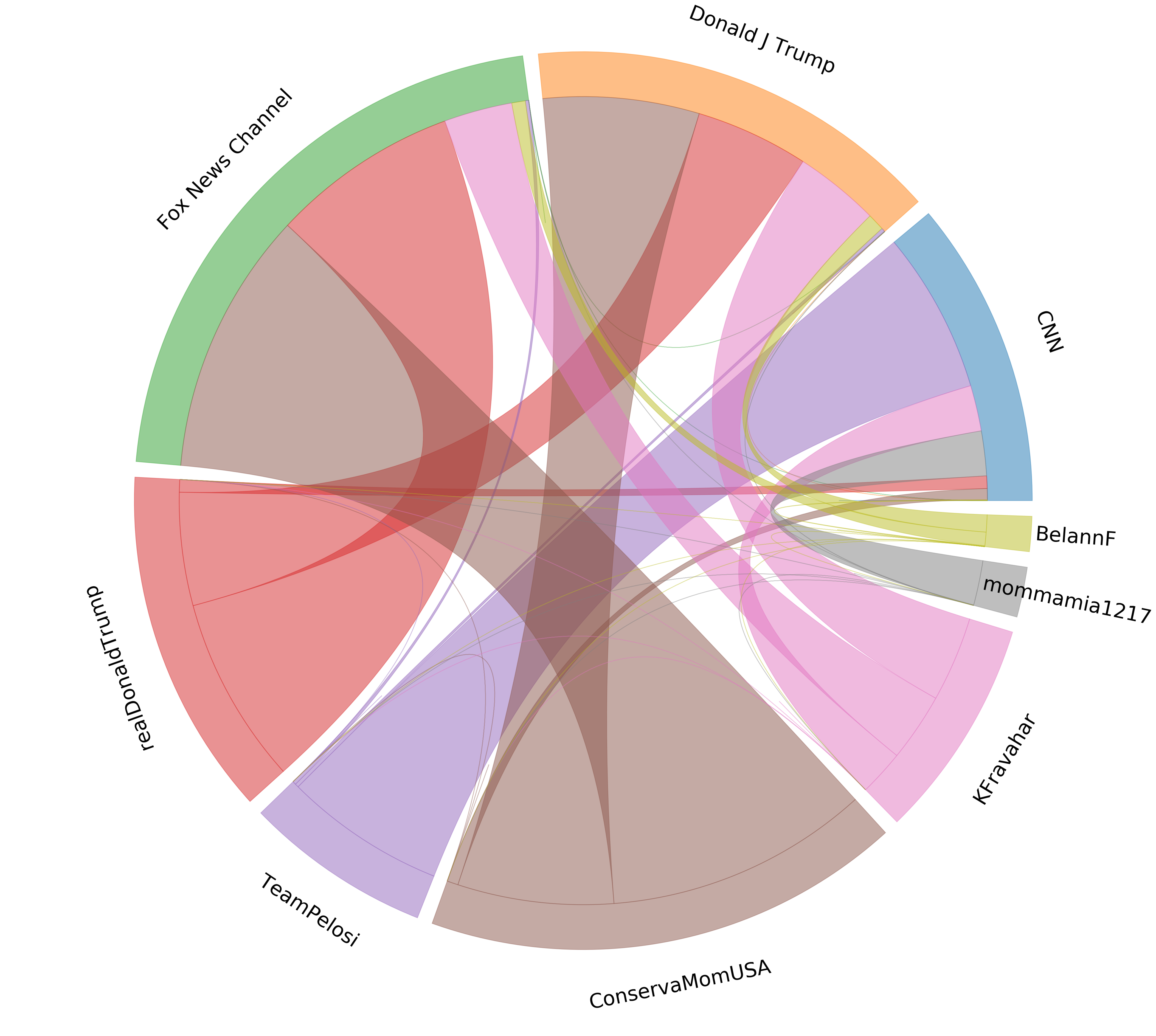}
\caption{Interactions between largest communities in YouTube and Twitter.}
\label{fig:yt-tw}
\end{figure}

Figure \ref{fig:yt-tw} shows the interactions between the 3 largest YouTube communities (top half) in the YouTube-comment graph (YT) and the 6 largest Twitter communities in the Retweet graph (RT). Each community is named after its highest PageRank member (or second highest, when more clear) in the corresponding graph. The size of each relation depicts the number of users in the RT community that tweeted video URLs from any channel in the YT community.

\section{Conclusions}
This work having obtained the tweets for the most popular hashtags regarding the
US elections 2020, as well as the the extracted unique YouTube videos, performs
an analysis regarding the volume of tweets and users, the identification of
entities and the correlation between the features of the YouTube videos. Next,
it demonstrates sentiment analysis on the Twitter corpus and the YouTube metadata
and shows that the positive sentiment is higher for Donald Trump in comparison
with Joe Biden. We identify how real world events trigger user discussions in social media around
that topic. Additionally, this study includes the evolution of the retweet
graph along six different time points in the dataset, from July to September
2020 and highlights the two main entities (’Biden’ and ’Trump’).

We are planning to perform sarcasm detection in this dataset, which is an important
step in the analysis of political content.

\begin{acks}
This document is the results of the research project co-funded by the
European Commission, project CONCORDIA, with grant number 830927
(EUROPEAN COMMISSION Directorate-General Communications Networks,
Content and Technology) and by the European Union and Greek national
funds through the Operational Program Competitiveness,
Entrepreneurship and Innovation, under the call
RESEARCH--CREATE--INNOVATE (project code:T1EDK-02857, and
T1EDK-01800).
\end{acks}

\appendix

\begin{table}[ht]
\begin{tabular}{|c|c|}
\hline
Hashtag & Tweets count \\
\hline
\#Trump2020& 2.930.633\\
\#Vote & 1.785.378\\
\#vote & 1.494.138 \\
\#Election2020 &  1.277.839\\
\#Biden  & 1.129.063\\
\#VoteBlueToSaveAmerica  & 802.793\\
\#trump2020 & 228.289 \\
\#VoteTrumpOut & 147.065\\
\#election2020 &  106.412\\
\#trump &  89.767\\
\#biden & 82.802\\
\#2020election & 41.794\\
\#November3rd & 34.299\\
\#NovemberIsComing & 32.680\\
\#donaldtrump & 28.265\\
\#MyPresident & 24.609\\
\#Elections\_2020 & 6.249\\
\#2020elections & 3.828\\
\#USElections & 3.580\\
\#bluewave2020 &  3.040\\
Total & 10.252.523\\
\hline
\end{tabular}
\caption{The list of all hashtags in our dataset.}
\label{table:allhashtags}
\end{table}

\section{List of all Hashtags}
\label{appendix}
In this section, we list all the retrieved hashtags on which we based our dataset along with the tweets count within which there were contained, in table \ref{table:allhashtags}.

\bibliographystyle{ACM-Reference-Format}











\end{document}